\documentclass[10pt,conference]{IEEEtran}
\def\BibTeX{{\rm B\kern-.05em{\sc i\kern-.025em b}\kern-.08em
    T\kern-.1667em\lower.7ex\hbox{E}\kern-.125emX}}

\usepackage{float}

\usepackage[numbers]{natbib}

\usepackage[utf8]{inputenc}

\usepackage{csquotes}

\usepackage{graphicx}
\graphicspath{{images/}}

\PassOptionsToPackage{hyphens}{url}\usepackage{hyperref}

\usepackage{array}

\usepackage{multirow}

\usepackage{xcolor}

\begin{document}

\title{
    Design Patterns for AI-based Systems: A Multivocal Literature Review and Pattern Repository
}

\author{
    \IEEEauthorblockN{Lukas Heiland}
    \IEEEauthorblockA{
        University of Stuttgart\\
        Institute of Software Engineering\\
        Stuttgart, Germany\\
        st151210@stud.uni-stuttgart.de
    }
    \and
    \IEEEauthorblockN{Marius Hauser}
    \IEEEauthorblockA{
        University of Stuttgart\\
        Institute of Software Engineering\\
        Stuttgart, Germany\\
       st171565@stud.uni-stuttgart.de
    }
    \and
    \IEEEauthorblockN{Justus Bogner}
    \IEEEauthorblockA{
        University of Stuttgart\\
        Institute of Software Engineering\\
        Stuttgart, Germany\\
        justus.bogner@iste.uni-stuttgart.de
    }
}

\maketitle

% \IEEEpeerreviewmaketitle

\begin{abstract}
Systems with artificial intelligence components, so-called AI-based systems, have gained considerable attention recently.
However, many organizations have issues with achieving production readiness with such systems.
As a means to improve certain software quality attributes and to address frequently occurring problems, \textit{design patterns} represent proven solution blueprints.
While new patterns for AI-based systems are emerging, existing patterns have also been adapted to this new context. 

The goal of this study is to provide an overview of design patterns for AI-based systems, both new and adapted ones.
We want to collect and categorize patterns, and make them accessible for researchers and practitioners.
To this end, we first performed a multivocal literature review (MLR) to collect design patterns used with AI-based systems.
We then integrated the created pattern collection into a web-based pattern repository to make the patterns browsable and easy to find.

As a result, we selected 51 resources (35 white and 16 gray ones), from which we extracted 70 unique patterns used for AI-based systems.
Among these are 34 new patterns and 36 traditional ones that have been adapted to this context.
Popular pattern categories include \textit{architecture} (25 patterns), \textit{deployment} (16), \textit{implementation} (9), or \textit{security \& safety} (9).
While some patterns with four or more mentions already seem established, the majority of patterns have only been mentioned once or twice (51 patterns).
Our results in this emerging field can be used by researchers as a foundation for follow-up studies and by practitioners to discover relevant patterns for informing the design of AI-based systems.
\end{abstract}


\section{Introduction}
Software systems relying on artificial intelligence (AI) have gained considerable attention in industry and academia~\cite{AiIndex2022}.
As such, AI-based systems are systems which contain one or several AI software components, with potentially different responsibilities~\cite{EU2018}.
However, a large percentage of AI projects ($\sim$70-80\%) is estimated to fail~\cite{Schmelzer2022,Gartner2018}, without making it to production or staying there in sustainable fashion for longer periods of time.

To tackle these problems, the discipline of AI Engineering was formed~\cite{Bosch2021}.
While AI engineering practices~\cite{Serban2022} can be one way to improve quality, another approach more focused on architecture and design are \textit{patterns}.
According to Gamma et al.~\cite{Gamma1994}, design patterns are proven, reusable solutions to frequently occurring problems.
Even though the impact of individual design patterns on software quality is sometimes controversial,
%~\cite{Khomh2008, Hegedus2012}
a recent systematic literature review by \citet{Wedyan2020} reports an overall positive impact of design patterns on software quality.

While new design patterns are slowly emerging for AI-based systems, it is currently difficult to get a recent overview of the pattern landscape in this space.
In 2019, \citet{Washizaki2022} conducted a first multivocal review and reported design patterns for machine learning (ML) systems.
The authors suggested repeating the study after several years, since the research topic would be actively researched and many new design patterns could establish after the study.

Our goal is therefore to provide an extensive, updated overview of design patterns used for AI-based systems, and to create an extendable pattern repository to document the results.
By surveying ML practitioners, Washizaki et al.~\cite{Washizaki2022} discovered that 31\% solve problems in an ad-hoc way and do not consciously use existing design patterns.
The effort for finding and using design patters should therefore be reduced, both to raise awareness and to allow the wide-spread usage of successful design patterns.

To address this, we conducted a multivocal literature review (MLR), using both scientific (white) and practitioner (gray) resources.
We especially wanted to find patterns published after the search window of \citet{Washizaki2022}, traditional patterns adapted to the AI context, and patterns not restricted to ML systems, i.e., ones applicable to AI-based systems in general.
This study will answer the following research questions:
\begin{itemize}
    \item \textbf{RQ1:} Which new design patterns have emerged for AI-based systems?
    \item \textbf{RQ2:} Which existing design patterns have been adapted to AI-based systems?
    \item \textbf{RQ3:} How can these patterns be consistently documented and made accessible for researchers and practitioners?
\end{itemize}

During the identification of patterns for AI-based systems, we consciously differentiated between \textit{new} patterns that emerged specifically for AI-based systems and existing, \textit{traditional} patterns from other fields that were adapted for AI-based systems.
In addition to finding and categorizing the patterns, a web-based pattern repository has been created to make the pattern collection easily browsable and accessible.

\section{Background and Related Work}
In this section, we first introduce important fundamentals, namely AI-based system and design patterns, and then discuss existing related work in the area.

\subsection{AI-based Systems}
AI-based systems are systems which include AI components~\cite{martinez2022software}.
They can be purely software-based, e.g., voice assistants, or AI can be embedded in the hardware, e.g., in Internet of Things applications.
These systems often learn by analyzing their environment, and then take actions based on the analysis, with the goal to express intelligent behavior.
While other subfields of AI are relevant for such system, machine learning (ML)~\cite{Jordan2015} is most frequently used.
Due to the presence of AI components, e.g., with embedded ML models, AI-based systems exhibit characteristics that make them distinct from other software applications, such as an increased importance of data quality and management, unclear abstraction boundaries for complex models, and challenges in the customization and reuse of AI components~\cite{bogner2021characterizing}.

\subsection{Design Patterns}
A design pattern is a proven and established solution to a recurring design problem that is documented in a technology-agnostic form~\cite{Gamma1994}.
A pattern generally consists at least of the four essential elements \textit{pattern name}, \textit{problem}, \textit{solution}, and \textit{consequences}.
As such, it is more a set of guidelines rather than a specific sequence of instructions.
This means that implementation details can vary between different applications of a pattern.
The idea of patterns originated from building architecture~\cite{alexander1977pattern}, but has been successfully adopted by the software engineering community, most notably by~\citet{Gamma1994}, the \enquote{The Gang of Four}.
Today, design patterns have emerged in almost every area of software engineering, including, but not limited to, cloud computing~\cite{Fehling2014}, microservices~\cite{taibi2018architectural}, and messaging~\cite{hohpe2004enterprise}.
Patterns codify expert knowledge and therefore enable conversations at a higher level of abstraction, leading to quicker design decisions~\cite{rising2010benefit}.

\subsection{Related Work}
While many systematic reviews have been conducted for general design patterns~\cite{BafandehMayvan2017} or patterns in other domains like microservices~\cite{taibi2018architectural} or IoT~\cite{Washizaki2020b}, there is only one study that collects design patterns for AI-based systems.
In 2019, \citet{Washizaki2022} conducted an MLR to identify software engineering design patterns in the field of machine learning, which was published in 2022.
They identified 15 design patterns for this kind of application, and additionally analyzed their perception and usage via a survey with 118 ML practitioners.
Their only used search engine for white literature was Engineering Village\footnote{\url{https://www.engineeringvillage.com}}, which provides access to 12 engineering databases.
Their gray literature search was performed using the Google search engine.
To find relevant resources within a broader definition of design patterns, they also included keywords such as \enquote{implementation pattern} and \enquote{architecture pattern}.
Their queries produced 80 resources, from which they included 38 resources after analysis.
From these, they extracted 33 patterns, but reduced the number to 15 after additional quality checks.
They classified the patterns into three different categories (\textit{topology}, \textit{programming}, and \textit{model operation}) and assigned nine quality attributes.
Among these are five system and software product quality attributes, namely performance efficiency, compatibility, security, maintainability, and portability, as well as four model and prediction quality attributes: model robustness, model explainability, prediction accuracy, and prediction fairness.
Lastly, the authors mention a lack of well-described patterns and missing standardization.

Since design patterns for ML systems is an active research topic, \citet{Washizaki2022} suggested conducting a comparable study again in the future to get an up-to-date list of design patterns.
The literature search in~\cite{Washizaki2022} was conducted in August 2019.
The main reason to apply design patterns is to quickly find solutions to recurring problems so that a designer can spend more time on thinking about creative solutions to problems not yet covered by patterns~\cite{Diaz2008}.
In a field where many organizations have issues with creating production-ready AI-based systems, having no knowledge of and no convenient overview of the pattern landscape can negatively impact many quality attributes.
So far, AI-related design patterns were mainly published in books or blog posts~\cite{Lakshmanan2020, Samaali2022, Shibui2021}, with only a few scientific papers~\cite{Washizaki2022, Sharma2019}.
This means there is no comprehensive collection of design patterns which is also easy to browse.
Our study aims to fill this gap by providing an up-to-date collection of design patterns for AI-based systems and an easily usable pattern repository.

\section{Study Design}
A multivocal literature review (MLR) is a special type of systematic literature review (SLR) that not only uses scientific white literature, but also practitioner resources, namely gray literature~\cite{Garousi2016}.
This provides the benefit that additional resources close to a practical context can be included when not many scientific publications exist for a new topic yet.
\citet{Garousi2016} argue that an industry-oriented discipline like software engineering should consider conducting MLRs in addition to SLRs, as gray literature would provide valuable input in emerging fields.
Resources categorized as white literature (WL) are academic peer-reviewed articles.
Gray literature (GL) are practitioner resources such as books, web pages, or white papers, all of which are not peer-reviewed.
According to the Luxembourg definition~\cite{encyInfo2009}, gray literature \enquote{is produced on all levels of government, academics, business and industry in print and electronic formats, but which is not controlled by commercial publishers, i.e., where publishing is not the primary activity of the producing body}.
One potential challenge with gray literature is to judge whether a resource contains reliable content.
To help with this, \citet{Garousi2019} proposed guidelines for working with GL and how to decide whether a resource is relevant and trustworthy.

\subsection{Process}
\begin{figure}[ht]
    \centering
    \includegraphics[width=\columnwidth]{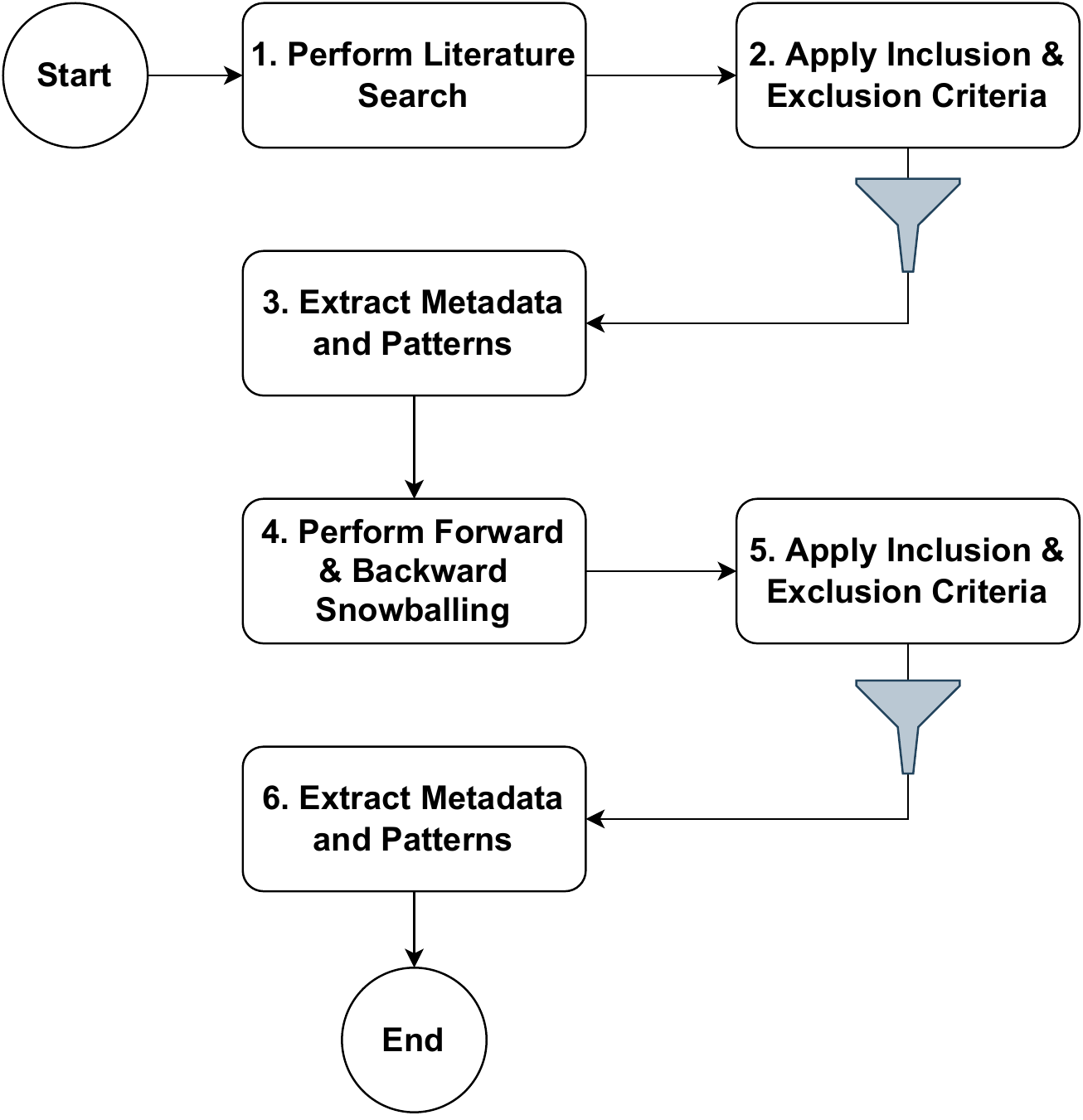}
    \caption{General process of the conducted MLR}
    \label{fig:searchStrat}
\end{figure}

In the following, we list the general steps of our review process.
This process is also visualized in Fig.~\ref{fig:searchStrat}.
More details for each step are presented in the upcoming subsections.

\begin{enumerate}
    \item Perform initial literature search using defined query strings. The work is split between two researchers.
    \item Apply inclusion and exclusion criteria. Selection of resources is again split between two researchers. Results from one person are reviewed by the other, applying the four-eyes principle. Disagreements are discussed and decisions are made together with all researchers.
    \item Perform pattern extraction on the included resources to ensure that all resources contain at least one design pattern for AI-based system according to the pattern quality criteria.
    The work was split between the first two authors. Each author extracted half of the found resources. Afterwards, the authors reviewed each other's extractions. In case of disagreement, the third author was included to make a decision.
    \item Perform one round of forward and backward snowballing using all included resources as start set.
    The identified relevant resources were split, with the first two authors each performing snowballing for one half and reviewing the other's results afterwards.
    \item Perform pattern extraction as in step 3.
\end{enumerate}

\subsection{Search Strategy}
For white literature, we selected the two databases IEEE Xplore and ACM Digital Library, plus the search engine Google Scholar.
Combined with snowballing, this provides broad coverage of the relevant literature.
For gray literature, we chose the Google search engine, as it is the most popular and encompassing provider.

Inspired by the keywords used by \citet{Washizaki2022}, we formed our own set of search terms to fit a broad definition of design patterns for our targeted systems.
After consulting newer resources on the topic known to us, we settled on the following key terms:
\begin{quote}
    \textit{software engineering, ai-based systems, artificial intelligence, machine learning, pattern, design, architecture, recipe}.
\end{quote}
We specifically excluded hits containing the term \textit{recognition} to avoid papers about pattern recognition.
Based on this, we formulated our basic query string as the following boolean combination of search terms:

\begin{quote}
    {\fontfamily{cmtt}\selectfont
        "software engineering" AND ("ai-based systems" OR "artificial intelligence" OR ai OR machine learning) AND (pattern OR design OR architecture OR recipe) AND NOT recognition
    }
\end{quote}

For GL, we split the string into several pairs because the Google search engine does not handle complex strings well, e.g., parentheses cannot be used.\footnote{\url{https://support.google.com/websearch/answer/2466433}}
Therefore, several queries of the following form were constructed:
\begin{quote}
    {\fontfamily{cmtt}\selectfont
        ”artificial intelligence” AND pattern AND NOT recognition \\
        ”artificial intelligence” AND design AND NOT recognition
    }
\end{quote}
For all these queries, we merged the results while excluding duplicates.
The full list of query strings can be found in our provided digital appendix.\footnote{\url{https://doi.org/10.5281/zenodo.7568062}}

To allow for collaboration during resource selection and later data extraction, the identified resources were collected in an online Excel spreadsheet.
Since not all used search engines can be accessed in the same way, different approaches were required to export and merge the data.
For Google Scholar, we used the tool \textit{Publish or Perish}\footnote{\url{https://harzing.com/resources/publish-or-perish}} to export 100 search results at once into an Excel spreadsheet.
Since Google Scholar cannot process complex queries well, the search keywords were split up into 20 separate queries (see appendix for all query strings).
For each of the 20 search queries, the first 100 search results were considered and extracted.
Both IEEE Xplore and ACM Digital Library provide export capabilities, which were used to extract the first 2,000 search results produced by the complex query string described above.
However, since both libraries only support the export of a small number of hits at once (ACM DL: 50, IEEE Xplore: 100), the export was still a bit of manual work.
All three exports also have slightly different metadata attributes, which we consolidated into a common format.
During this process, we also removed duplicates.

Since Google Search does not provide support to export search results, the results were extracted using a Python script based on BeautifulSoup\footnote{\url{https://www.crummy.com/software/BeautifulSoup}}.
The script is an adaption of Google2Csv\footnote{\url{https://github.com/psalias2006/Google2Csv}}.
Google search results only provide the page title, page link, and an excerpt of the page content.
Additional metadata like publication year or authors needed to be filled in manually during data extraction.
Just like Google Scholar, Google Search cannot process complex queries.
Therefore, the same 20 pairs were used as query strings, with the first 100 search results per query string being exported.
All in all, we therefore considered a maximum of 2,000 search hits for each of the four sources in the selection process.

\subsection{Resource and Pattern Selection}
To be included in our MLR, a resource had to satisfy certain criteria.
To apply these criteria, we used \textit{adaptive reading depth}~\cite{Petersen2008}, i.e., a researcher read as much (or little) of a resource as they needed to decide about inclusion or exclusion with reasonable certainty.
First, the full-text of a resource had to be written in English and needed to be available to us.
Second, a resource had to mention or describe at least one pattern in the context of AI-based systems.
These criteria applied to both WL and GL.
Additionally, we had a quality criterion for including individual patterns from a resource, which was especially important for the non-peer-reviewed GL.
A pattern needed to be described with at least the following attributes to be included, i.e., simply mentioning a pattern name was \textit{not} enough for inclusion:

\begin{itemize}
	\item Pattern name
	\item Motivation (describing the problem to be solved)
	\item Solution (describing how to solve the problem)
\end{itemize}

These criteria are inspired by \citet{Gamma1994}, but also by \citet{Fehling2014}, who constructed a pattern language for cloud computing.
Their extensive pattern template contains the following attributes: \textit{pattern name}, \textit{intent}, \textit{icon}, \textit{driving question}, \textit{context}, \textit{solution}, \textit{sketch}, \textit{result}, \textit{variations}, \textit{related patterns}, and \textit{known uses}.
However, in the emerging field of AI-based systems, many patterns are not yet formalized as extensively and rigorously, which means that attributes like \textit{sketch} or \textit{related patterns} are not always available.
We therefore focused on the bare minimum set of pattern attributes that provide a base level of usefulness.

\subsection{Data Extraction}
After resource selection, we extracted data for all included resources.
Based on the exported search metadata, we defined a common extraction sheet for both WL and GL, which comprised the following attributes:

\begin{itemize}
	\item Resource ID (given by us)
	\item Resource title
	\item Authors
	\item DOI / URL
	\item Year
	\item Resource type (WL / GL)
\end{itemize}

For WL, most of these attributes were already provided by the metadata export.
Extracting the individual patterns described in a resource was a more complex process, for which we first had to derive a uniform data structure to which the identified patterns can be converted.
Based on \citet{Gamma1994} and \citet{Fehling2014}, we decided on a set of attributes that would work well for our context.
Attributes like \textit{sample code}~\cite{Gamma1994} assume that the pattern can be easily illustrated with source code.
Since most found patterns are of an architectural or processes-related nature, such attributes do not make a lot of sense in our case.
Additionally, many emerging AI patterns are only documented sparsely.
To reduce the overall number of empty attributes in our collection, we sometimes combined multiple related attributes into a single, more generic one.
For example, \textit{implementation}, \textit{sample code}, and \textit{known uses} were aggregated into an attribute called \textit{examples}.
We applied a similar aggregation to \textit{intent}, \textit{problem}, and \textit{motivation}.
The final pattern template, which we also use as the basic data structure for the pattern repository, consists of the following attributes:

\begin{itemize}
    \item id: ID of the pattern (given by us)
    \item name: the name of the pattern
    \item aka: potential aliases of the pattern
    \item categories: a list of pattern categories
    \item motivation: the reason to use the pattern, i.e., which problem does it solve?
    \item solution: the solution to the problem
    \item consequences: the consequences that arise from using the pattern (both positive and negative ones)
    \item examples: illustrations or known uses of the pattern
    \item related patterns: names of related pattern
    \item resources: list of resource IDs that mention the pattern
\end{itemize}

Some resources, mostly secondary studies such as \cite{Washizaki2022, Washizaki22020, Washizaki2020, Lu2022}, provided a fairly orderly view of design patterns and their attributes.
This allowed a comparably fast extraction of patterns.
However, most primary studies and gray literature resources, such as \cite{Samaali2022,Sinha1992,Kum2020}, only presented patterns in plain text, which required more time to judge if sufficient information was provided and to extract them.

\subsection{Snowballing}
To extend the initial search results, we performed one round of forward and backward snowballing following \citet{Wohlin2014}.
For backward snowballing, the reference list of a resource was used.
Some gray literature resources also mentioned references and were therefore considered for backward snowballing as well.
For forward snowballing, the \enquote{cited by} features of Google Scholar and IEEE Xplore have been used.
ACM DL also provides a \enquote{cited by} feature, but no resource published by ACM was included in the snowballing start set.
In all cases, IEEE Xplore provided a subset of the resources provided by Google Scholar, i.e., incorporating IEEE Xplore for citation search did not bring any benefits.
We only performed one snowballing iteration because the additionally included resources did not yield many new patterns and mostly supplied additional occurrences for already found patterns.
For gray literature, no forward snowballing was carried out because there is no convenient way to analyze links pointing to a web resource.

\section{MLR Results}
\begin{figure}[ht]
    \centering
    \includegraphics[width=0.9\columnwidth]{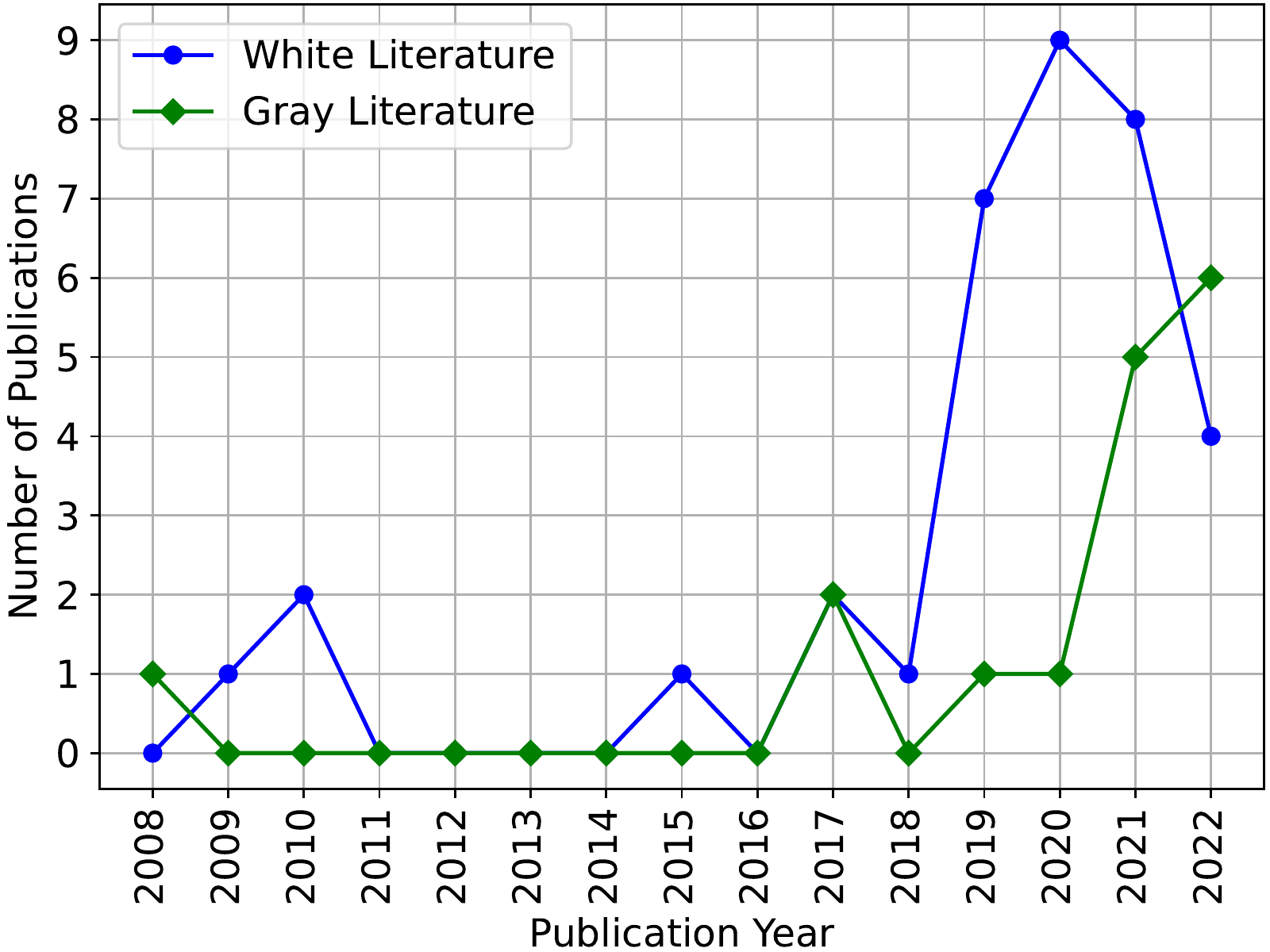}
    \caption{Publication year distribution of gray and white literature}
    \label{fig:pubyears}
\end{figure}

Compared to Washizaki et al.~\cite{Washizaki2022}, our study revealed many additional resources and patterns for AI-based systems.
The initial search queries resulted in 2,961 unique resources, of which 1,304 were white and 1,657 were gray literature.
After applying inclusion and exclusion criteria, 29 resources were selected as relevant, i.e., they mentioned or introduced at least one pattern for AI-based systems with sufficient details.
Afterwards, snowballing was performed.
The 29 included resources were cited by 76 other resources, from which, excluding duplicates, 7 were classified as relevant during forward snowballing.
Backward snowballing revealed 269 references, from which, excluding duplicates, 15 were classified as relevant.
For the most part, however, snowballing produced additional mentions of patterns that had already been identified in the initial query-based search.
One reason for this was that we already identified secondary studies~\cite{Washizaki2022} or publications referencing multiple patterns~\cite{Lu2022} in the first round.
A particularly effective snowballing resource was a preliminary study~\cite{Washizaki2019} of the related work of \citet{Washizaki2022}.
It led to 12 relevant resources using backward snowballing and 8 relevant resources using forward snowballing.
This can be explained with the considerable attention their study got and the fact that they conducted the only secondary study in this research area. 
In total, we identified 51 relevant resources (35 for WL, 16 for GL) and extracted 70 unique patterns from them.
Fig.~\ref{fig:pubyears} shows the number of publications per year, separated into gray and white literature.

The period during which the resources were published ranges from 2008 to 2022.
The related work of \citet{Washizaki2022} included literature until 2019-08-19.
Most of the resources found in our MLR were published after this date, which explains the greater number of found patterns for AI-based systems.
We see a strong increase in publications in the years 2019 and following. 
From the 16 GL resources, all but one were published between 2017 and 2022.
The majority of the 35 found WL resources were also published after 2017.

\begin{figure}[ht]
    \centering
    \includegraphics[width=0.6\columnwidth]{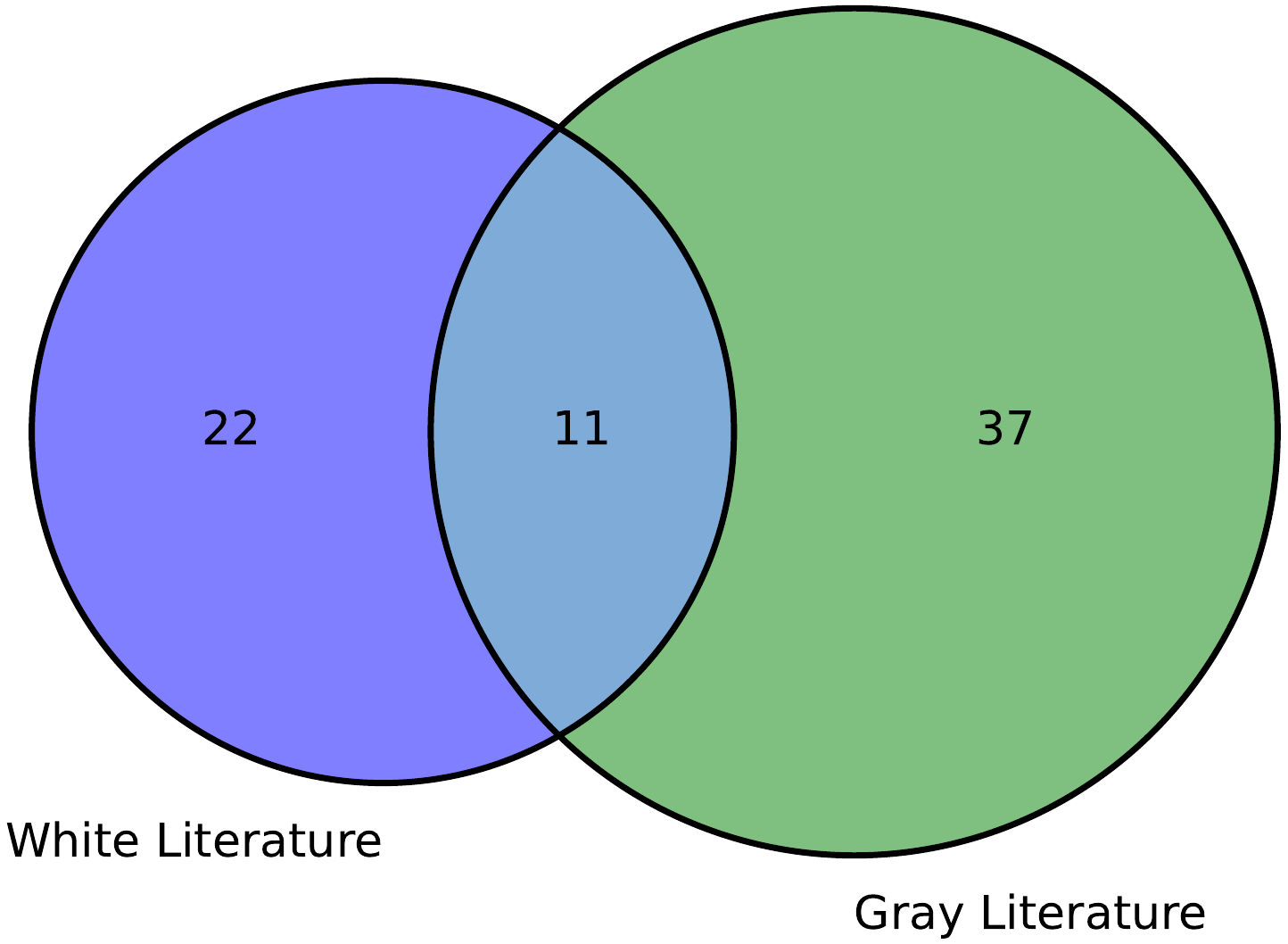}
    \caption{Mentions of the 70 patterns differentiated by gray and white literature}
    \label{fig:vennWLGL}
\end{figure}

Fig.~\ref{fig:vennWLGL} shows a Venn diagram of the origin of the 70 patterns in our review, i.e., how many were mentioned by gray, white, or both gray and white literature.
37 patterns received mentions exclusively from gray literature resources.
22 pattern received only references from white literature, and 11 patterns received mentions from both.
This indicates that many patterns for AI-based systems are mentioned in gray literature, but that they have not yet been adopted by publications in white literature.
Since most of the gray literature resources have been published since 2017, some pattern may be covered in white literature in the coming years.
All in all, the consideration of gray literature yielded 37 new patterns and 11 mentions of patterns that are also discussed in academia.

\begin{figure*}[ht]
    \centering
    \includegraphics[width=0.7\textwidth]{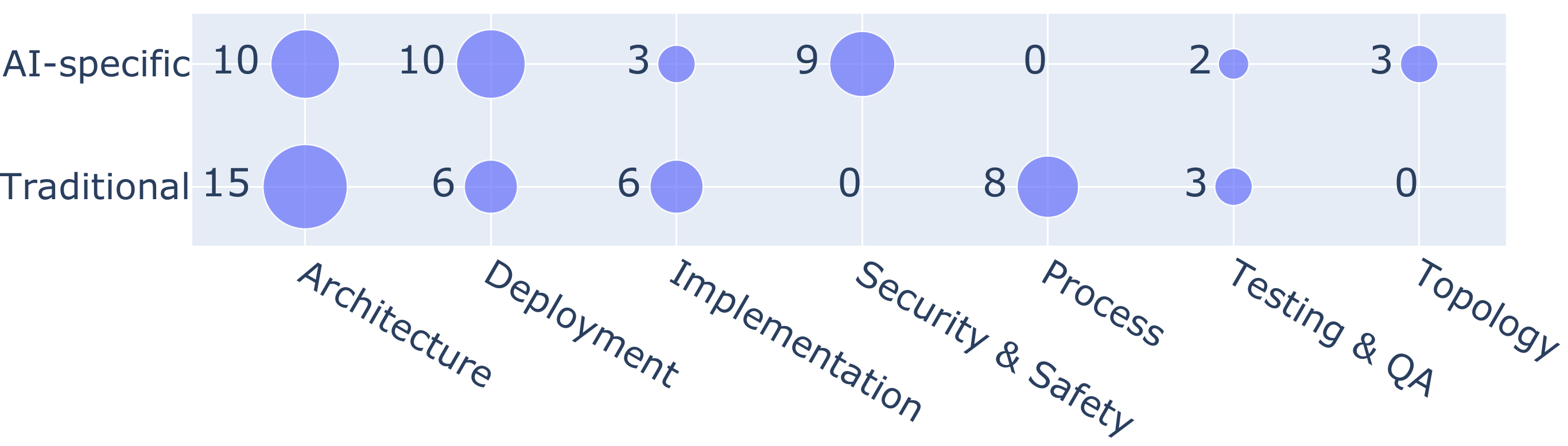}
    \caption{Comparison of the number of adapted traditional (36) and new AI-specific (34) patterns; a pattern can have multiple categories}
    \label{fig:patternsByCategory}
\end{figure*}

Our review identified a total of 70 unique patterns for AI-based systems.
These patterns were extracted from 51 resources, from which 35 are academic and 16 non-academic.
Of the 70 identified patterns, 36 were \textit{traditional} patterns, i.e., existing patterns that were adapted to the AI context.
The remaining 34 were new patterns that specifically emerged for AI-based systems ($\sim$49\%).
In addition to \textit{traditional} vs. \textit{new}, we divided the found patterns into the following categories:

\begin{itemize}
    \item architecture (25 patterns)
    \item deployment (16 patterns)
    \item implementation (9 patterns)
    \item security \& safety (9 patterns)
    \item process (8 patterns)
    \item testing \& quality assurance (5 patterns)
    \item topology (3 patterns)
\end{itemize}

These categories emerged as a mixture of existing categorizations, refined with adaptations based on our understanding of the holistic data set.
Like \citet{Washizaki2022}, we also considered categorizing the patterns based on improved quality attributes from ISO 25010.
However, we abandoned this idea after discovering that the majority of identified patterns did not explicitly describe this.
A few patterns were assigned to multiple categories because they were too extensive in scope.

\begin{figure}[H]
    \centering
    \includegraphics[width=\columnwidth]{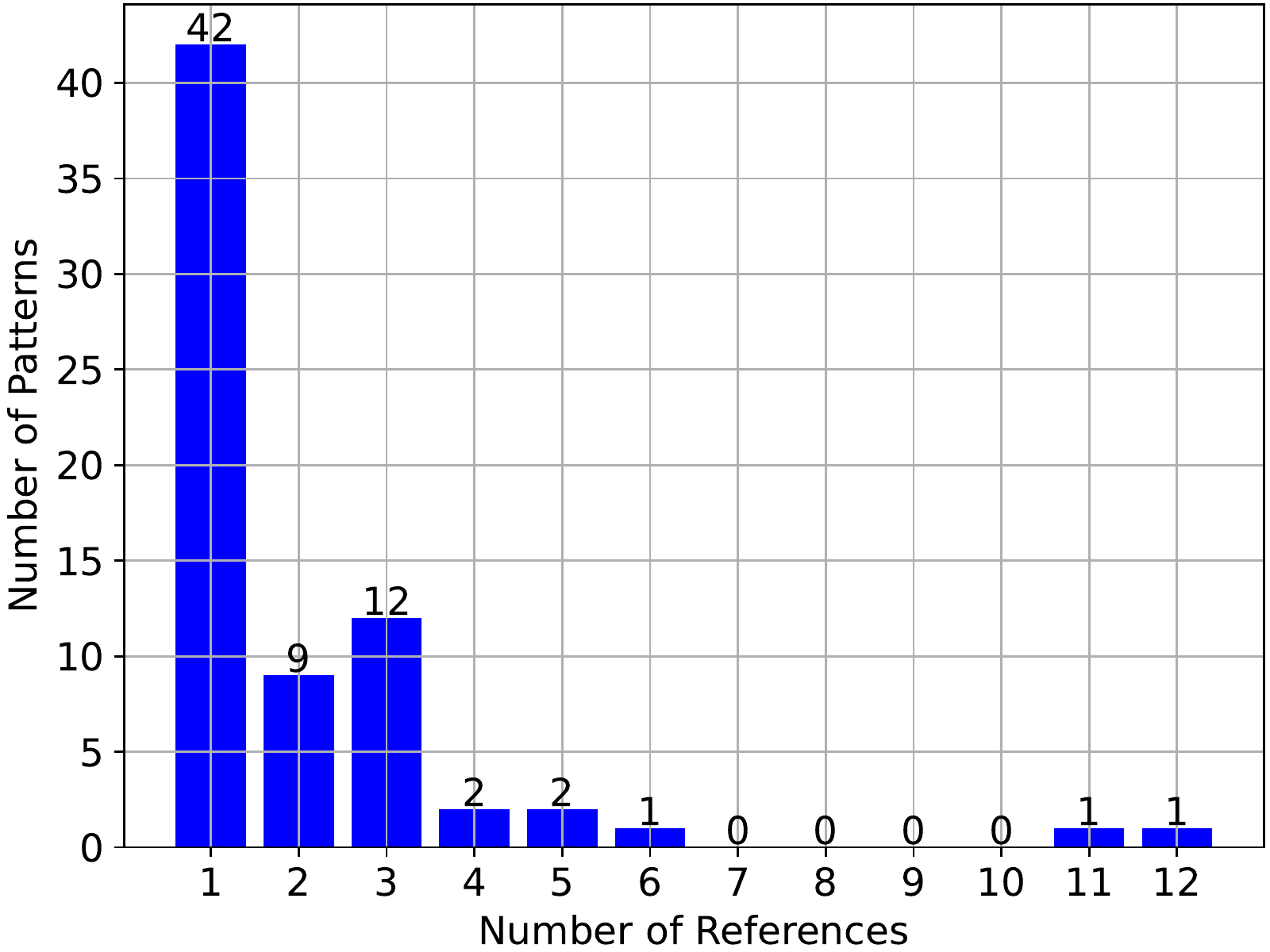}
    \caption{Number of patterns per received number of references by resources}
    \label{fig:resourceCnt}
\end{figure}

Fig.~\ref{fig:patternsByCategory} shows the number of patterns in each category, split into AI-specific and traditional.
One interesting discovery is that most patterns only received one or two mentions from resources.
Fig.~\ref{fig:resourceCnt} shows how many patterns received a certain number of references.
42 of 70 patterns were mentioned in only a single resource, and 9 were mentioned twice.
Very popular patterns mentioned in five or more resources were uncommon.
Consulting practitioners like \citet{Washizaki2022} could reveal whether patterns that received only one or two references are truly unknown among practitioners.
However, among all found patterns, some stand out due to their high number of references in white and gray literature.
An excerpt of these most mentioned individual patterns is presented in Table~\ref{tab:patterns}.

\subsection{New Design Patterns for AI-based Systems (RQ1)}
Roughly half of the identified patterns (34 of 70, $\sim$49\%) were new ones that had emerged specifically for AI-based systems.
The most popular categories here were \textit{architecture} (10 patterns), \textit{deployment} (10), and \textit{security} \& safety (9).
While the first one was also frequently assigned to \textit{traditional} patterns (15), the latter two were more interesting, as together with \textit{topology} (3), they were very focused on new patterns.
For example, not a single \textit{traditional} pattern was assigned to \textit{security \& safety} or \textit{topology}.
In the following, we use the described template to give some examples of new AI-specific patterns from these categories.
\\\\
\textit{Name}: \textbf{Workflow Pipeline}\\
\textit{aka}: -\\
\textit{Motivation}: Creating an end-to-end reproducible training and deployment pipeline for a machine learning component is difficult.
Data science notebooks can run a whole pipeline, but they do not scale.\\
\textit{Solution}: Make each pipeline step a separate containerized service.
Services are orchestrated and chained together to form pipelines that can be run via REST API calls.\\
\textit{Consequences}: The portability, scalability, and maintainability of the individual pipeline steps is improved, at the cost of an overall more complex solution.\\
\textit{Examples}: Presented at AWS Blog\footnote{\url{https://aws.amazon.com/blogs/machine-learning/build-mlops-workflows-with-amazon-sagemaker-projects-gitlab-and-gitlab-pipelines}}\\
\textit{Related}: -\\
\textit{Categories}: architecture, topology\\
\textit{Resources}: \cite{Tkachuk2020,Granlund2021,Changyao2020,Mo2021,Lakshmanan2020,Sharma2019,Ippolito2022,Yan2022,Saby2021,Rios2008,Zinkevich2022}
\\\\
\textit{Name}: \textbf{Two-Phase Predictions}\\
\textit{aka}: Cascade Predictions, Multiple Stage Prediction Pattern\\
\textit{Motivation}: Executing large, complex models can be time-consuming and costly, especially if lightweight clients like mobile or IoT devices are involved.\\
\textit{Solution}: Split the prediction into two phases.
A simple, fast model is executed first on the client.
Afterwards, a large, complex model is optionally executed in the cloud for deeper insights.\\
\textit{Consequences}: Prediction response time is reduced for some cases.
The number of large, expensive predictions is reduced.
The client has a fall-back model when there is no Internet connection.\\
\textit{Examples}: Voice activation in AI assistants like Alexa or Google Assistant\\
\textit{Related}: -\\
\textit{Categories}: deployment\\
\textit{Resources}: \cite{Lakshmanan2020,Saby2021}
\\\\
\textit{Name}: \textbf{Encapsulating ML Models within Rule-based Safeguards}\\
\textit{aka}: -\\
\textit{Motivation}: It is impossible to guarantee the correctness of ML model predictions, so they should not be directly used for safety- or security-related functions.
Furthermore, ML models can be unstable and vulnerable to adversarial attacks, data noise, and drift.\\
\textit{Solution}: Introduce a deterministic, rule-based mechanism that decides what to do with the prediction results, e.g., based on additional quality checks.\\
\textit{Consequences}: Reduced risk for negative impacts of incorrect predictions, but a more complex architecture.\\
\textit{Examples}: -\\
\textit{Related}: -\\
\textit{Categories}: security \& safety\\
\textit{Resources}: \cite{Takeuchi2021,Washizaki2019,Washizaki22020,Washizaki2020,Washizaki2022}
\\\\
\textit{Name}: \textbf{AI Pipelines}\\
\textit{aka}: Sequential Decomposition\\
\textit{Motivation}: Complex prediction or synthesis use cases are often difficult to accomplish with a single AI tool or model.\\
\textit{Solution}: Divide the problem into smaller consecutive steps, then combine several existing AI tools or custom models into an inference-time AI pipeline, where each specialized tool or model is responsible for a single step.\\
\textit{Consequences}: More tools and models need to be integrated, but the provided is result is of higher quality.
Each step can be optimized individually.\\
\textit{Examples}: Typical computer vision inference pipelines as described in~\cite{Mo2021}\\
\textit{Related}: Pipes and Filters\\
\textit{Categories}: architecture, topology\\
\textit{Resources}: \cite{Tkachuk2020,Mo2021,Yan2022}
\\\\
\textit{Name}: \textbf{Ethics Credentials}\\
\textit{aka}: Verifiable Ethical Credentials\\
\textit{Motivation}: Responsible AI requirements are either omitted or mostly stated as high-level objectives, and not specified explicitly in a verifiable way as expected system outputs.
Because of this, users may trust an AI system less, or even refrain from using it.\\
\textit{Solution}: Provide verifiable ethics credentials for your AI system or component.
Using publicly accessible and trusted data infrastructure, the credentials can be verified as proof of ethical compliance.
Additionally, users may also have to verify their credentials before getting access to the AI system.\\
\textit{Consequences}: Trust and system acceptance increases, and awareness of ethical issues is raised.
However, a trusted public data infrastructure is needed, and credentials need to be maintained and potentially refreshed from time to time.\\
\textit{Examples}: -\\
\textit{Related}: -\\
\textit{Categories}: security \& safety\\
\textit{Resources}: \cite{Lu2022,Lu2022a}

\begin{table*}
    \centering
    \caption{Excerpt of most frequently mentioned patterns}
    \label{tab:patterns}
    \small
    \begin{tabular}{
        p{0.15\textwidth}
        p{0.3\textwidth}
        p{0.3\textwidth}
        p{0.08\textwidth}
        p{0.08\textwidth}
    }
    \textbf{Pattern Name} &
      \textbf{Problem / Motivation} &
      \textbf{Solution / Application} &
      \textbf{Categories} &
      \textbf{Resources} \\
      \hline
      \hline
    Multi-Layer Pattern &
      An application comprises several groups of subtasks, each of which is at a different level of abstraction. &
      Divide the application into different layers. Each layer, consisting of sub-modules, can be independently designed to feed into foreign prototypes. Each layer has an input and corresponding output to the succeeding layer. The succeeding return is used as a feed to the next layers for further processing. Every layer communicates only with its direct neighbor &
      architecture, traditional 
      & \cite{Sharma2019,Zhou2020,Bi2021,Wang2021,Yokoyama2019,Smith2017,Savolainen2009,Washizaki2019,Yokoyama2019,Smith2017,Washizaki22020,Khomh2020}\\
    Workflow Pipeline &
      Creating an end-to-end reproducible training and deployment pipeline for a machine learning component is difficult. Data science notebooks can run a whole pipeline, but they do not scale. &
      Make each pipeline step a separate containerized service. Services are orchestrated and chained together to form pipelines that can be run via REST API calls. &
      architecture, topology 
      & \cite{Tkachuk2020,Granlund2021,Changyao2020,Mo2021,Lakshmanan2020,Sharma2019,Ippolito2022,Yan2022,Saby2021,Rios2008,Zinkevich2022}\\
    Lambda Architecture Pattern &
      Different components in a system have different performance requirements that need to be satisfied, e.g., some are focused on throughput and others on response time. &
      Group the components based on their latency requirements into three layers: (1) batch layer ingests and stores large amounts of data (2) speed layer processes updates to the data in low-latency (3) serving layer provides precalculated results in a low-latency fashion &
      architecture, traditional
      & \cite{Haindl2022,John2017,Washizaki2019,Khomh2020,Washizaki22020,Washizaki2022}\\
      Distinguish Business Logic from ML Model &
      Machine Learning (ML) systems are complex because their ML components must be (re)trained regularly and have an intrinsic non-deterministic behavior. Similar to other systems, the business requirements for these systems as well as ML algorithms change over time. &
      Define clear APIs between traditional and ML components. Place business and ML components with different responsibilities into three layers. Divide data flows into three. &
      architecture
      & \cite{Washizaki2019,Ao2010,Yokoyama2019,Khomh2020,Washizaki2022}\\
      Encapsulating ML Models within Rule-based Safeguards &
    Since prediction by Machine Learning (ML) models is often difficult to guarantee correctness, ML models cannot be used directly for security-related functions. Furthermore, ML models are unstable and vulnerable to adversarial attacks, data noise, and drift.
    & Introduce a rule-based mechanism and provide the rule-matched results as outputs for the specific inputs 
    & security \& safety
    & \cite{Takeuchi2021,Washizaki2019,Washizaki22020,Khomh2020,Washizaki2022} \\
    Microservice Vertical Pattern 
    & When to use: when you need to run several inferences in order or when you have several inferences and they have dependencies
    &  The pattern deploys prediction models in separate servers or containers as services. You execute prediction requests from top to bottom synchronously, and gather the results to respond to the client.
    & architecture
    & \cite{Shibui2021,Smith2017,Khomh2020,Washizaki2022} \\
    Microservice Horizontal Pattern 
    & When to use: when the workflow can execute multiple predictions in parallel or when you want to integrate prediction results at last. Required to run several predictions to one request
    & Multiple are deployed models in parallel. You can send one request to the models at once to acquire multiple predictions, or an integrated prediction.
    & architecture
    & \cite{Shibui2021,Smith2017,Khomh2020,Washizaki2022} \\
    AI Pipeline
    & Complex prediction or synthesis use cases are often difficult to accomplish with a single AI tool or model.
    &  Divide the problem into smaller consecutive steps, then combine several existing AI tools or custom models into an inference-time AI pipeline, where each specialized tool or model is responsible for a single step.
    & architecture, topology
    & \cite{Tkachuk2020,Mo2021,Yan2022} \\
    \hline
    \hline
    \end{tabular}
\end{table*}

\subsection{Adaptation of Traditional Design Patterns for AI (RQ2)}
We categorized existing patterns that were adapted to the AI context as \textit{traditional} patterns.
From the 70 patterns, 36 were such traditional ones ($\sim$51\%).
For example, among them is the well-known Client-Server pattern, originally published in 1992~\cite{Sinha1992}.
\citet{Sharma2019} mentioned the adaption of this pattern for AI-based systems in 2019.
For these patterns, the most prominent categories were \textit{architecture} (15 patterns), \textit{process} (8), and \textit{implementation} (6).
All these categories also have more traditional than new patterns, even though the difference is more pronounced for the latter two.
For example, \textit{process} contains exclusively traditional patterns.
This may suggest that challenges in these areas may not differ so much between AI and non-AI systems that a large number of fundamentally new patterns are required.
Many \textit{implementation} patterns are also adaptations of the original \enquote{Gang of Four} design patterns to an AI context, e.g., \textit{Strategy}, \textit{Adapter}, or \textit{Decorator}.
Regarding \textit{architecture}, \citet{Sculley2015} stated that ML components only make up a small portion of ML systems, i.e., many ML challenges are not limited to their ML components.
This might explain why so many traditional architecture patterns have been adapted to this context.
Some traditional patterns like \textit{Multi-Layer Pattern} or \textit{Lambda Architecture} even received a large number of mentions.
In the following, we present some examples.
\\\\
\textit{Name:} \textbf{Multi-Layer Pattern}\\
\textit{aka:} Separation of Concerns, Multi-Tiered Architecture\\
\textit{Motivation:} An application comprises several groups of subtasks, each of which is at a different level of abstraction.\\
\textit{Solution:} Divide the application into different layers.
Each layer consists of submodules, and can be independently designed to feed into foreign prototypes.
It has an input and corresponding output to the succeeding layer.
The succeeding return is used as a feed to the next layer for further processing.
Every layer communicates only with its direct neighbor.\\
\textit{Consequences:} This pattern enables inference of results at the individual steps.
Extra components can be added or modified to meet the computational requirements, giving it high flexibility.\\
\textit{Examples:} self-learning student platform described in~\cite{Zhou2020}\\
\textit{Related:} -\\
\textit{Categories:} architecture, traditional\\
\textit{Resources:} \cite{Sharma2019,Zhou2020,Bi2021,Wang2021,Yokoyama2019,Smith2017,Savolainen2009,Washizaki2019,Yokoyama2019,Smith2017,Washizaki22020,Khomh2020}
\\\\
\textit{Name:} \textbf{Continuous Integration and Deployment}\\
\textit{aka:} -\\
\textit{Motivation:} Reduce the risk of releasing broken applications.\\
\textit{Solution:} Always build with unit and component tests and deploy with verification tests using code that itself is under version control.
After you commit to a development branch, the system deploys to a development environment. Once all end-to-end and manual smoke testing is complete, a manual action deploys to production.\\
\textit{Consequences:} -\\
\textit{Examples:} \texttt{ease.ml/ci} presented in~\cite{Renggli2019}\\
\textit{Related:} Continued Model Evaluation, End-To-End Tests\\
\textit{Categories:} process, traditional\\
\textit{Resources:} \cite{Gustafson2022,Serban2022,Renggli2019}
\\\\
\textit{Name:} \textbf{Strategy Pattern}\\
\textit{aka:} -\\
\textit{Motivation:} How can an ML model that performs a task in a given context be flexibly changed?\\
\textit{Solution:} Define an interface (strategy) that different models implement.
The context will call the methods exposed by the interface, and the implemented models will behave differently based on the contextual data.\\
\textit{Consequences:} Switching models or achieving flexibility in model behavior is easier, but code complexity is increased.\\
\textit{Examples:} XGBoost's custom objective functions, Hugging Face's pipeline interface\\
\textit{Related:} -\\
\textit{Categories:} implementation, traditional\\
\textit{Resources:} \cite{Take2021,Samaali2022,Yan2022}
\\\\
\textit{Name:} \textbf{Deploy Canary Model}\\
\textit{aka:} -\\
\textit{Motivation:} You trained a new model with assumed better prediction quality, but it's not certain if this will carry over to production.
Additionally, there could be other quality issues with the new model that should not affect all users in production at once.\\
\textit{Solution:} Deploy the new model in addition to the existing ones and route a small number of requests to it to evaluate its performance.
If this test is successful, all existing models can be replaced.
If not, the new model needs to be improved.\\
\textit{Consequences:} Only a small number of users are subjected to potential bugs or low-quality predictions.
Additional serving and monitoring infrastructure is required.\\
\textit{Examples:} -\\
\textit{Related:} -\\
\textit{Categories:} deployment, traditional\\
\textit{Resources:} \cite{Washizaki22020,Washizaki2020,Khomh2020}
\\\\
\textit{Name:} \textbf{Batch Serving}\\
\textit{aka:} -\\
\textit{Motivation:} Predictions need to be carried out asynchronously over large volumes of data (contrary to predictions for small, individual requests), e.g., generating personalized playlists every week. This is only applicable if there is no need for (near-)real-time predictions.\\
\textit{Solution:} Use distributed data processing infrastructure (e.g, based on MapReduce) to asynchronously carry out complex ML inference tasks on a large number of computing nodes.
The individual predictions are aggregated back into a single result.\\
\textit{Consequences:} Positive: You can manage server resources flexibly and in strict accordance with demand. You may rerun the job in case of error. There is no requirement for high availability in your server system. Negative: You need a job management server. This pattern depends on the ability to split a task across multiple workers.\\
\textit{Examples:} -\\
\textit{Related:} Stateless Serving Function\\
\textit{Categories:} deployment, traditional\\
\textit{Resources:} \cite{Lakshmanan2020,Shibui2021,Saby2021}

\section{Web-based Pattern Repository}
To make the findings of our MLR easily accessible and browsable, we developed a web application, in which the identified patterns can be viewed.
The application and its user interface is inspired by an existing repository for service-based antipatterns~\cite{Bogner2019-CSEQUDOS}.
Our application was built as a single page application using the web framework Vue.js\footnote{\url{https://vuejs.org}}, with the component library \textit{Vuetify}\footnote{\url{https://vuetifyjs.com}} for UI building blocks.
The repository is hosted on GitHub\footnote{\url{https://github.com/SWE4AI/ai-patterns}}, with the web app being accessible on GitHub Pages\footnote{\url{https://swe4ai.github.io/ai-patterns}} via a web browser.

\begin{figure*}[ht]
    \centering
    \includegraphics[width=\textwidth]{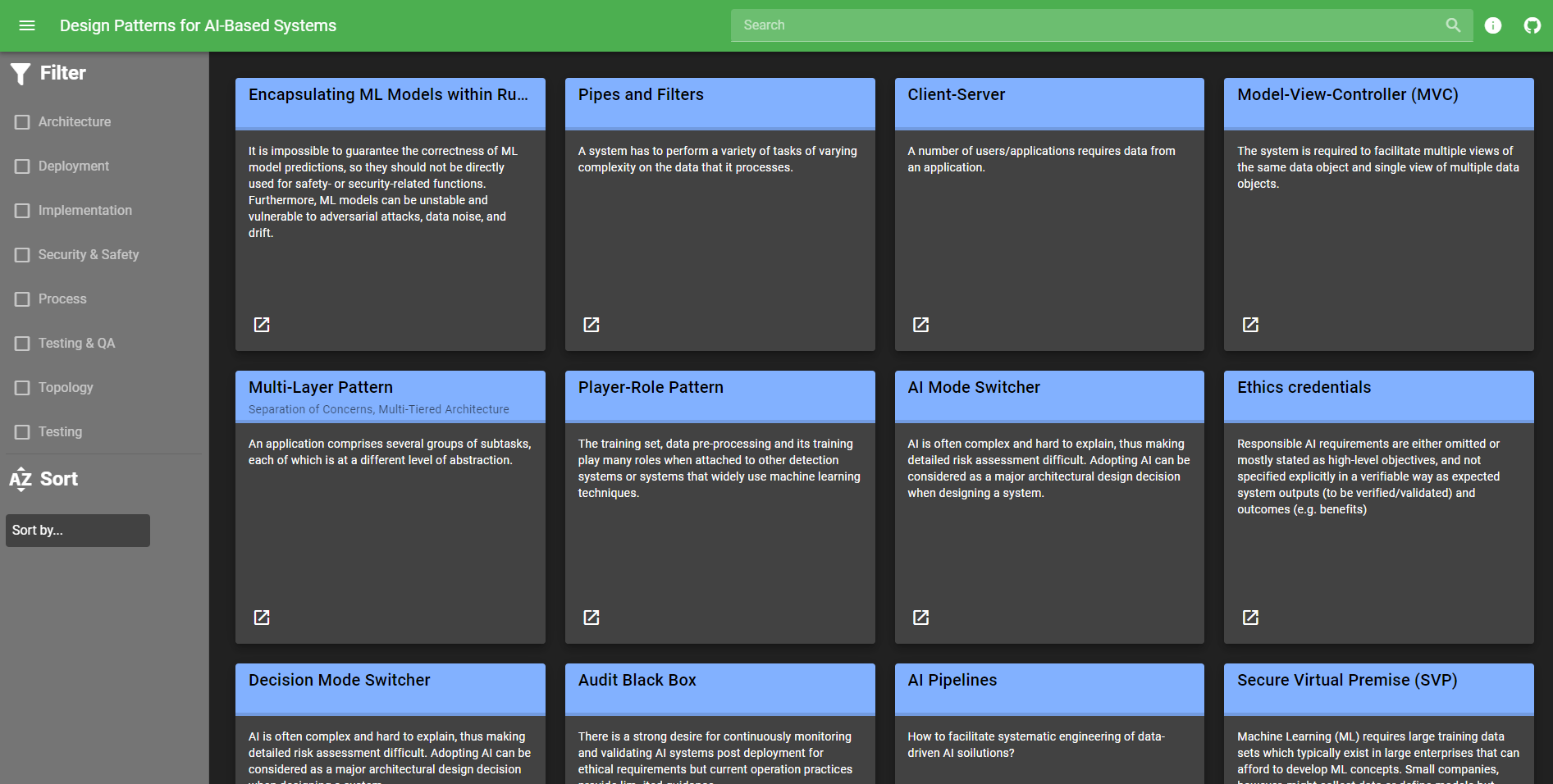}
    \caption{Screenshot of the main view of the pattern repository web application}
    \label{fig:patternRepositoryWebApp}
\end{figure*}

As mentioned, we synthesized a template based on existing pattern collections to document our identified patterns.
For the repository, this template needed to be expressed in a machine-readable format, for which we chose JSON.
All patterns in the existing spreadsheet were converted into this JSON structure by using the tool CSV2JSON\footnote{\url{https://csvjson.com/csv2json}}.
For increased flexibility, each pattern has its own JSON file that is part of the repository.
Resources are documented in a separate JSON files, and are referenced via their respective IDs in the pattern JSON files.
This makes the repository easily changeable and extendable.
Refining the data for one pattern only changes a single, fairly small JSON file, and, once a new pattern has been identified, a JSON file with the required attributes can be created and committed to the repository, all without changing the source code.
Moreover, this allows to use collaborative features of GitHub like issues and pull requests to continuously refine the pattern repository within the research community.

The web application features a dashboard-like interface, of which we present a screenshot in Fig.~\ref{fig:patternRepositoryWebApp}.
The main part of the view dynamically displays the pattern collection, organized as cards. 
Each card shows the pattern name, an alias (if existing), and a brief description of the problem that is solved by this pattern.
To see more attributes, every pattern card can be expanded via a pop-up into a detail view, showing all information gathered from the MLR.
Citations for the literature which mentions a specific paper can be exported to the user's clipboard.
Via the sidebar, the pattern collection can be filtered by the identified categories and sorted according to pattern name and number of references.
Additionally, a search bar is included at the top of the user interface, which can be used for free-text filtering.

\section{Discussion}
In the following, we try to provide an interpretation of our results.
One interesting finding in our review is the roughly equal distribution between existing, \textit{traditional} patterns that were adapted to an AI context on the one hand and \textit{new}, AI-specific patterns on the other hand.
However, the two favor slightly different categories, with traditional patterns being, e.g., more prominent in \textit{process} and \textit{implementation}, and new patterns in \textit{security \& safety} and \textit{deployment}.
This is related to the discussion whether we really need new SE practices for AI-based systems.
One faction of researchers and practitioners argues that, yes, we indeed need many new ones, while the other side argues that using existing practices in slightly adapted form would be mostly sufficient.
Our MLR results paint a nuanced picture of this discussion.
Yes, many identified patterns are clear adaptations or even non-adapted applications of existing ones, e.g., \textit{Deploy Canary Model}~\cite{Washizaki2020} or \textit{Discard Proof of Concept Code}~\cite{Washizaki2022}.
However, we also found many patterns that specifically emerged in the AI context and are of less or no value elsewhere, e.g., \textit{Ethics Credentials}~\cite{Lu2022},
\textit{Two-Phase Predictions}~\cite{Lakshmanan2020}, or \textit{Parameter-Server Abstraction}~\cite{Washizaki2022}.
So, while existing SE practices and patterns clearly hold value for AI-based systems, we believe there is definitely the need for several new ones.

Another interesting finding was the prominence of the \textit{architecture} pattern category, which was the largest one for both \textit{new} (10) and \textit{traditional} (15) patterns.
It also included many of the most frequently mentioned patterns (see Table~\ref{tab:patterns}), adding to the perceived importance of reusing architectural knowledge in this field.
One explanation for the popularity of this category might be the danger of hidden dependencies and unexpected component interactions that can occur in AI-based systems.
\citet{Sculley2015} have described this phenomenon as the CACE principle: \enquote{Changing Anything Changes Everything}.
Relying on architectural abstractions of different forms to isolate change, e.g., encapsulation via custom interfaces, may therefore be a promising solution, which is reflected in the abundance of architectural patterns.

In this review, the inclusion of gray literature greatly influenced the results.
37 patterns were exclusively described in GL resources, and 11 additional ones also received GL mentions.
This highlights the potential of non-peer-reviewed practitioner resources in emerging fields.
Interestingly, GL resources tended to mention more patterns per resource than WL, and also discussed a broader range of patterns.
This led to the discovery of 3 patterns per GL resource (48 patterns identified via 16 resources) and only $\sim$0.94 patterns per WL resource (33 patterns identified via 35 resources).
However, this pattern density in the GL resources comes at a price.
First, patterns exclusively described in GL only averaged 1.2 mentions per pattern.
This is less than for patterns only described in WL (2.1 mentions / pattern) and considerably less for patterns described by both (4.7 mentions / pattern).
From the 51 patterns that were only mentioned once or twice, 35 exclusively occurred in GL ($\sim$69\%).
Since patterns are not \enquote{created}, but instead \enquote{mined} from analyzing existing working solutions in industrial practice (\enquote{Rule of Three}\footnote{\url{https://wiki.c2.com/?RuleOfThree}}), some of these instances might not even be considered as true patterns (yet).
Additionally, several of these patterns had only fairly shallow descriptions that did not follow a consistent template, which led to, e.g., very short or even missing attribute descriptions in our pattern collection.
This definitely decreases their value, especially for software professionals looking to implement them for their AI-based systems.
For completeness, analysis, and future extension, we nonetheless included them into our repository.
To ignore them, the ability to sort by the number of references can serve as an imperfect quality filter.

\section{Threats to Validity}
Even though we followed established guidelines for MLRs and adhered to a rigorous review protocol, several limitations have to be mentioned for the results of this study.
Regarding \textbf{internal validity}, slight errors or inconsistencies might have happened during the study selection, data extraction, or data analysis of our review.
We used the four-eyes principle as much as possible, but also split up tasks in the data extraction and snowballing phase.
Since the results of these tasks are sometimes open to interpretation, slight variations might exist with a different team of researchers.
Additionally, categorizing patterns is also an inherently subjective process.
While three researchers were involved in this activity and discussed any difference of opinion, it is likely that other researchers would produce a slightly different categorization.
For example, a more conservative categorization could have led to fewer \textit{new} patterns and more \textit{traditional} ones.
For transparency and replicability, we publish our review artifacts on Zenodo\footnote{\url{https://doi.org/10.5281/zenodo.7568062}}.

Regarding \textbf{external validity}, a common threat with systematic reviews is if enough relevant resources have been identified for the extraction and analysis.
If the percentage of the population is too small, the results might not be generalizable.
Our study identified more resources and patterns than any previous works, and we believe that we achieved a reasonably broad coverage of the AI pattern landscape.
Nonetheless, it is possible that a few resources and their associated patterns remained unidentified.
While we used snowballing in both directions, we only performed one level of it.
Still, we believe the risk that important, frequently used patterns have been missed to be very small.

Furthermore, some of our results could be less valuable for industry because many pattern descriptions are incomplete.
Limited information about the solution or consequences could discourage practitioners from using a pattern.
Additionally, a lack of concrete implementation examples might mean that some patterns may be hard to apply in practice, even if practitioners would consider them useful.

\section{Conclusion}
To provide an overview of design patterns for AI-based systems, we conducted a multivocal literature review.
In total, we extracted and documented 70 patterns from 51 resources, thereby finding many relevant patterns for AI-based systems that have not been mentioned in existing secondary studies.
Roughly half of the identified patterns (36) were \textit{traditional} ones that had been adapted to an AI context, with the other half being new, AI-specific patterns (34).
For convenient browsing and filtering, we created a public web-based pattern repository\footnote{\url{https://github.com/SWE4AI/ai-patterns}}, which may benefit both practitioners and researchers interested in design pattern for AI-based systems.

Future work could refine the existing pattern collection, e.g., by improving the extracted pattern attributes, adding more relationships between patterns, or adding new patterns to the collection.
Additionally, creating new forms of visualization, e.g., a pattern map as proposed by \citet{Washizaki2022} that makes it easy to grasp the relationships among patterns, may be helpful.
The pattern repository could also be complemented with antipatterns for AI-based systems, i.e., frequently occurring suboptimal solutions with mostly negative consequences.
Lastly, the importance or effects of the patterns could be analyzed via practitioner surveys or controlled experiments.

\section*{Acknowledgment}
This research was partially funded by the Ministry of Science, Research, and the Arts (MWK) Baden-Württemberg, Germany, within the Artificial Intelligence Software Academy (AISA).

\bibliographystyle{IEEEtranN}
\bibliography{references}

% Generated by IEEEtranN.bst, version: 1.14 (2015/08/26)
\begin{thebibliography}{61}
\providecommand{\natexlab}[1]{#1}
\providecommand{\url}[1]{#1}
\csname url@samestyle\endcsname
\providecommand{\newblock}{\relax}
\providecommand{\bibinfo}[2]{#2}
\providecommand{\BIBentrySTDinterwordspacing}{\spaceskip=0pt\relax}
\providecommand{\BIBentryALTinterwordstretchfactor}{4}
\providecommand{\BIBentryALTinterwordspacing}{\spaceskip=\fontdimen2\font plus
\BIBentryALTinterwordstretchfactor\fontdimen3\font minus
  \fontdimen4\font\relax}
\providecommand{\BIBforeignlanguage}[2]{{%
\expandafter\ifx\csname l@#1\endcsname\relax
\typeout{** WARNING: IEEEtranN.bst: No hyphenation pattern has been}%
\typeout{** loaded for the language `#1'. Using the pattern for}%
\typeout{** the default language instead.}%
\else
\language=\csname l@#1\endcsname
\fi
#2}}
\providecommand{\BIBdecl}{\relax}
\BIBdecl

\bibitem[for Security and via AI~Index~(2022)(2022)]{AiIndex2022}
\BIBentryALTinterwordspacing
C.~for Security and E.~T. via AI~Index~(2022). (2022) Annual number of
  scholarly publications on artificial intelligence. [Online]. Available:
  \url{https://ourworldindata.org/grapher/number-ai-publications?country=~OWID_WRL}
\BIBentrySTDinterwordspacing

\bibitem[EU(2018)]{EU2018}
\BIBentryALTinterwordspacing
EU. (2018) A definition of ai: Main capabilities and scientific disciplines.
  [Online]. Available:
  \url{https://ec.europa.eu/digital-single-market/en/high-level-expert-group-artificial-intelligence}
\BIBentrySTDinterwordspacing

\bibitem[Schmelzer(2022)]{Schmelzer2022}
\BIBentryALTinterwordspacing
R.~Schmelzer, ``The one practice that is separating the ai successes from the
  failures,'' \emph{Forbes}, 2022. [Online]. Available:
  \url{https://www.forbes.com/sites/cognitiveworld/2022/08/14/the-one-practice-that-is-separating-the-ai-successes-from-the-failures/?sh=10b107e217cb}
\BIBentrySTDinterwordspacing

\bibitem[van~der Meulen and McCall(2018)]{Gartner2018}
\BIBentryALTinterwordspacing
R.~van~der Meulen and T.~McCall. (2018) Gartner says nearly half of cios are
  planning to deploy artificial intelligence. [Online]. Available:
  \url{https://www.gartner.com/en/newsroom/press-releases/2018-02-13-gartner-says-nearly-half-of-cios-are-planning-to-deploy-artificial-intelligence}
\BIBentrySTDinterwordspacing

\bibitem[Bosch et~al.(2021)Bosch, Olsson, and Crnkovic]{Bosch2021}
J.~Bosch, H.~H. Olsson, and I.~Crnkovic, ``Engineering {{AI Systems}}: {{A
  Research Agenda}},'' in \emph{Advances in {{Systems Analysis}}, {{Software
  Engineering}}, and {{High Performance Computing}}}, A.~K. Luhach and A.~El{\c
  c}i, Eds.\hskip 1em plus 0.5em minus 0.4em\relax {IGI Global}, 2021, pp.
  1--19.

\bibitem[Serban and Visser(2022)]{Serban2022}
A.~Serban and J.~Visser, ``Adapting {{Software Architectures}} to {{Machine
  Learning Challenges}},'' in \emph{2022 {{IEEE International Conference}} on
  {{Software Analysis}}, {{Evolution}} and {{Reengineering}}
  ({{SANER}})}.\hskip 1em plus 0.5em minus 0.4em\relax {IEEE}, 2022.

\bibitem[Gamma et~al.(1994)Gamma, Helm, Johnson, and Vlissides]{Gamma1994}
E.~Gamma, R.~Helm, R.~Johnson, and J.~Vlissides, \emph{Design patterns}.\hskip
  1em plus 0.5em minus 0.4em\relax Addison Wesley, 10 1994.

\bibitem[Wedyan and Abufakher(2020)]{Wedyan2020}
F.~Wedyan and S.~Abufakher, ``Impact of design patterns on software quality: A
  systematic literature review,'' \emph{IET Software}, vol.~14, pp. 1--17, 2
  2020.

\bibitem[Washizaki et~al.(2022)Washizaki, Khomh, Gueheneuc, Takeuchi, Natori,
  Doi, and Okuda]{Washizaki2022}
H.~Washizaki, F.~Khomh, Y.~G. Gueheneuc, H.~Takeuchi, N.~Natori, T.~Doi, and
  S.~Okuda, ``Software-engineering design patterns for machine learning
  applications,'' \emph{Computer}, vol.~55, pp. 30--39, 3 2022.

\bibitem[Mart{\'\i}nez-Fern{\'a}ndez et~al.(2022)Mart{\'\i}nez-Fern{\'a}ndez,
  Bogner, Franch, Oriol, Siebert, Trendowicz, Vollmer, and
  Wagner]{martinez2022software}
S.~Mart{\'\i}nez-Fern{\'a}ndez, J.~Bogner, X.~Franch, M.~Oriol, J.~Siebert,
  A.~Trendowicz, A.~M. Vollmer, and S.~Wagner, ``Software engineering for
  ai-based systems: a survey,'' \emph{ACM Transactions on Software Engineering
  and Methodology (TOSEM)}, vol.~31, no.~2, pp. 1--59, 2022.

\bibitem[Jordan and Mitchell(2015)]{Jordan2015}
M.~I. Jordan and T.~M. Mitchell, ``Machine learning: {{Trends}}, perspectives,
  and prospects,'' \emph{Science}, vol. 349, no. 6245, pp. 255--260, Jul. 2015.

\bibitem[Bogner et~al.(2021)Bogner, Verdecchia, and
  Gerostathopoulos]{bogner2021characterizing}
J.~Bogner, R.~Verdecchia, and I.~Gerostathopoulos, ``Characterizing technical
  debt and antipatterns in ai-based systems: A systematic mapping study,'' in
  \emph{2021 IEEE/ACM International Conference on Technical Debt (TechDebt)},
  2021, pp. 64--73.

\bibitem[Alexander(1977)]{alexander1977pattern}
C.~Alexander, \emph{A pattern language: towns, buildings, construction}.\hskip
  1em plus 0.5em minus 0.4em\relax Oxford university press, 1977.

\bibitem[Fehling et~al.(2014)Fehling, Leymann, Retter, Schupeck, and
  Arbitter]{Fehling2014}
C.~Fehling, F.~Leymann, R.~Retter, W.~Schupeck, and P.~Arbitter, \emph{Cloud
  {{Computing Patterns}}}.\hskip 1em plus 0.5em minus 0.4em\relax {Vienna}:
  {Springer}, 2014.

\bibitem[Taibi et~al.(2018)Taibi, Lenarduzzi, and Pahl]{taibi2018architectural}
D.~Taibi, V.~Lenarduzzi, and C.~Pahl, ``Architectural patterns for
  microservices: a systematic mapping study,'' in \emph{8th International
  Conference on Cloud Computing and Services Science}.\hskip 1em plus 0.5em
  minus 0.4em\relax SciTePress, 2018.

\bibitem[Hohpe and Woolf(2004)]{hohpe2004enterprise}
G.~Hohpe and B.~Woolf, \emph{Enterprise integration patterns: Designing,
  building, and deploying messaging solutions}.\hskip 1em plus 0.5em minus
  0.4em\relax Addison-Wesley Professional, 2004.

\bibitem[Rising(2010)]{rising2010benefit}
L.~Rising, ``The benefit of patterns,'' \emph{IEEE software}, vol.~27, no.~5,
  pp. 15--17, 2010.

\bibitem[Bafandeh~Mayvan et~al.(2017)Bafandeh~Mayvan, Rasoolzadegan, and
  Ghavidel~Yazdi]{BafandehMayvan2017}
B.~Bafandeh~Mayvan, A.~Rasoolzadegan, and Z.~Ghavidel~Yazdi, ``The state of the
  art on design patterns: {{A}} systematic mapping of the literature,''
  \emph{Journal of Systems and Software}, vol. 125, 2017.

\bibitem[Washizaki et~al.(2020{\natexlab{a}})Washizaki, Ogata, Hazeyama, Okubo,
  Fernandez, and Yoshioka]{Washizaki2020b}
H.~Washizaki, S.~Ogata, A.~Hazeyama, T.~Okubo, E.~B. Fernandez, and
  N.~Yoshioka, ``Landscape of architecture and design patterns for iot
  systems,'' \emph{IEEE Internet of Things Journal}, vol.~7, no.~10, pp.
  10\,091--10\,101, 2020.

\bibitem[Díaz et~al.(2008)Díaz, Aedo, and Rosson]{Diaz2008}
P.~Díaz, I.~Aedo, and M.~B. Rosson, ``Visual representation of web design
  patterns for end-users,'' \emph{Proceedings of the Workshop on Advanced
  Visual Interfaces}, 2008.

\bibitem[Lakshmanan et~al.(2020)Lakshmanan, Robinson, and Munn]{Lakshmanan2020}
V.~Lakshmanan, S.~Robinson, and M.~Munn, \emph{Machine learning design
  patterns}.\hskip 1em plus 0.5em minus 0.4em\relax O'Reilly Media, 10 2020.

\bibitem[Samaali(2022)]{Samaali2022}
\BIBentryALTinterwordspacing
T.~Samaali. (2022) Understand machine learning through 7 software design
  patterns. [Online]. Available:
  \url{https://towardsdatascience.com/understand-machine-learning-through-7-software-design-patterns-a03572f4e695}
\BIBentrySTDinterwordspacing

\bibitem[Shibui et~al.(2021)Shibui, Byeon, Seo, and Jin]{Shibui2021}
\BIBentryALTinterwordspacing
Y.~Shibui, S.~Y. Byeon, J.~Seo, and D.~Jin. (2021) System design patterns for
  machine learning. [Online]. Available:
  \url{https://github.com/mercari/ml-system-design-pattern}
\BIBentrySTDinterwordspacing

\bibitem[Sharma and Davuluri(2019)]{Sharma2019}
R.~Sharma and K.~Davuluri, ``Design patterns for machine learning
  applications,'' \emph{Proceedings of the 3rd International Conference on
  Computing Methodologies and Communication, ICCMC 2019}, pp. 818--821, 3 2019.

\bibitem[Garousi et~al.(2016)Garousi, Felderer, and Mäntylä]{Garousi2016}
V.~Garousi, M.~Felderer, and M.~V. Mäntylä, ``The need for multivocal
  literature reviews in software engineering: Complementing systematic
  literature reviews with grey literature,'' \emph{ACM International Conference
  Proceeding Series}, vol. 01-03-June, 6 2016.

\bibitem[McDonald and Levine-Clark(2017)]{encyInfo2009}
J.~D. McDonald and M.~Levine-Clark, \emph{Encyclopedia of library and
  information sciences}.\hskip 1em plus 0.5em minus 0.4em\relax CRC Press,
  2017.

\bibitem[Garousi et~al.(2019)Garousi, Felderer, and Mäntylä]{Garousi2019}
V.~Garousi, M.~Felderer, and M.~V. Mäntylä, ``Guidelines for including grey
  literature and conducting multivocal literature reviews in software
  engineering,'' \emph{Information and Software Technology}, vol. 106, pp.
  101--121, 2 2019.

\bibitem[Petersen et~al.(2008)Petersen, Feldt, Mujtaba, and
  Mattsson]{Petersen2008}
K.~Petersen, R.~Feldt, S.~Mujtaba, and M.~Mattsson, ``Systematic mapping
  studies in software engineering,'' \emph{12th International Conference on
  Evaluation and Assessment in Software Engineering}, vol.~17, 2008.

\bibitem[Washizaki et~al.(2020{\natexlab{b}})Washizaki, Uchida, Khomh, and
  Gu{\'e}h{\'e}neuc]{Washizaki22020}
H.~Washizaki, H.~Uchida, F.~Khomh, and Y.-G. Gu{\'e}h{\'e}neuc, ``Machine
  learning architecture and design patterns,'' \emph{Preprint}, 2020.

\bibitem[Washizaki et~al.(2020{\natexlab{c}})Washizaki, Takeuchi, Khomh,
  Natori, Doi, and Okuda]{Washizaki2020}
H.~Washizaki, H.~Takeuchi, F.~Khomh, N.~Natori, T.~Doi, and S.~Okuda,
  ``Practitioners' insights on machine-learning software engineering design
  patterns: A preliminary study,'' in \emph{IEEE International Conference on
  Software Maintenance and Evolution (ICSME)}, 2020.

\bibitem[Lu et~al.(2022{\natexlab{a}})Lu, Zhu, Xu, Whittle, Douglas, and
  Sanderson]{Lu2022}
Q.~Lu, L.~Zhu, X.~Xu, J.~Whittle, D.~Douglas, and C.~Sanderson, ``Software
  engineering for responsible ai: An empirical study and operationalised
  patterns,'' in \emph{44th International Conference on Software Engineering:
  Software Engineering in Practice}.\hskip 1em plus 0.5em minus 0.4em\relax
  IEEE, 2022.

\bibitem[Sinha(1992)]{Sinha1992}
A.~Sinha, ``Client-server computing,'' \emph{Communications of the ACM},
  vol.~35, pp. 77--98, 1 1992.

\bibitem[Kum et~al.(2020)Kum, Kim, Siracusa, and Moon]{Kum2020}
S.~Kum, Y.~Kim, D.~Siracusa, and J.~Moon, ``Artificial intelligence service
  architecture for edge device,'' \emph{IEEE International Conference on
  Consumer Electronics - Berlin, ICCE-Berlin}, vol. 2020-November, 11 2020.

\bibitem[Wohlin(2014)]{Wohlin2014}
C.~Wohlin, ``Guidelines for snowballing in systematic literature studies and a
  replication in software engineering,'' in \emph{18th International Conference
  on Evaluation and Assessment in Software Engineering}, 2014.

\bibitem[Washizaki et~al.(2019)Washizaki, Uchida, Khomh, and
  Guéhéneuc]{Washizaki2019}
H.~Washizaki, H.~Uchida, F.~Khomh, and Y.~G. Guéhéneuc, ``Studying software
  engineering patterns for designing machine learning systems,'' \emph{10th
  International Workshop on Empirical Software Engineering in Practice
  (IWESEP)}, 2019.

\bibitem[Tkachuk et~al.(2020)Tkachuk, Ilie, and Tutschku]{Tkachuk2020}
R.~V. Tkachuk, D.~Ilie, and K.~Tutschku, ``Towards a secure proxy-based
  architecture for collaborative ai engineering,'' \emph{8th International
  Symposium on Computing and Networking Workshops (CANDARW)}, 2020.

\bibitem[Granlund et~al.(2021)Granlund, Kopponen, Stirbu, Myllyaho, and
  Mikkonen]{Granlund2021}
T.~Granlund, A.~Kopponen, V.~Stirbu, L.~Myllyaho, and T.~Mikkonen, ``Mlops
  challenges in multi-organization setup: Experiences from two real-world
  cases,'' \emph{2021 IEEE/ACM 1st Workshop on AI Engineering - Software
  Engineering for AI, WAIN 2021}, 5 2021.

\bibitem[Chen(2020)]{Changyao2020}
\BIBentryALTinterwordspacing
C.~Chen. (2020) Machine learning design patterns: Reproducibility - pain is
  inevitable. suffering is optional. [Online]. Available:
  \url{https://changyaochen.github.io/ML-design-pattern-1/}
\BIBentrySTDinterwordspacing

\bibitem[Mo et~al.(2021)Mo, Oakes, and Galarnyk]{Mo2021}
\BIBentryALTinterwordspacing
S.~Mo, E.~Oakes, and M.~Galarnyk. (2021) Anyscale - serving ml models in
  production: Common patterns. [Online]. Available:
  \url{https://www.anyscale.com/blog/serving-ml-models-in-production-common-patterns}
\BIBentrySTDinterwordspacing

\bibitem[Ippolito(2022)]{Ippolito2022}
\BIBentryALTinterwordspacing
P.~P. Ippolito. (2022) Design patterns in machine learning for mlops | towards
  data science. [Online]. Available:
  \url{https://towardsdatascience.com/design-patterns-in-machine-learning-for-mlops-a3f63f745ce4}
\BIBentrySTDinterwordspacing

\bibitem[Yan(2022)]{Yan2022}
\BIBentryALTinterwordspacing
E.~Yan. (2022) Design patterns in machine learning code and systems. [Online].
  Available: \url{https://eugeneyan.com/writing/design-patterns/}
\BIBentrySTDinterwordspacing

\bibitem[Saby(2021)]{Saby2021}
\BIBentryALTinterwordspacing
N.~Saby. (2021) Machine learning in real life - machine learning design
  patterns. [Online]. Available:
  \url{https://mlinreallife.github.io/posts/ml-design-patterns/}
\BIBentrySTDinterwordspacing

\bibitem[Rios(2008)]{Rios2008}
\BIBentryALTinterwordspacing
G.~Rios. (2008) Patterns (and anti-patterns) for developing machine learning
  systems. [Online]. Available:
  \url{https://www.usenix.org/legacy/event/sysml08/tech/rios_talk.pdf}
\BIBentrySTDinterwordspacing

\bibitem[Zinkevich(2022)]{Zinkevich2022}
\BIBentryALTinterwordspacing
M.~Zinkevich. (2022) Rules of machine learning. [Online]. Available:
  \url{https://developers.google.com/machine-learning/guides/rules-of-ml}
\BIBentrySTDinterwordspacing

\bibitem[Takeuchi et~al.(2021)Takeuchi, Doi, Washizaki, Okuda, and
  Yoshioka]{Takeuchi2021}
H.~Takeuchi, T.~Doi, H.~Washizaki, S.~Okuda, and N.~Yoshioka, ``Enterprise
  architecture based representation of architecture and design patterns for
  machine learning systems,'' \emph{IEEE International Enterprise Distributed
  Object Computing Workshop, EDOCW}, 2021.

\bibitem[Lu et~al.(2022{\natexlab{b}})Lu, Zhu, Xu, and Whittle]{Lu2022a}
Q.~Lu, L.~Zhu, X.~Xu, and J.~Whittle, ``Responsible-ai-by-design: a pattern
  collection for designing responsible ai systems,'' \emph{arXiv}, 2022.

\bibitem[Zhou(2020)]{Zhou2020}
J.~Zhou, ``Design of ai-based self-learning platform for college english
  listening,'' in \emph{2nd International Conference on Machine Learning, Big
  Data and Business Intelligence (MLBDBI)}.\hskip 1em plus 0.5em minus
  0.4em\relax IEEE, 2020.

\bibitem[Bi et~al.(2021)Bi, Zhang, Yang, Jin, and Zhang]{Bi2021}
J.~Bi, G.~Zhang, C.~Yang, L.~Jin, and W.~Zhang, ``Architecture design of
  typical target detection and tracking system in battlefield environment,'' in
  \emph{3rd International Conference on Machine Learning, Big Data and Business
  Intelligence (MLBDBI)}.\hskip 1em plus 0.5em minus 0.4em\relax IEEE, 2021.

\bibitem[Wang(2021)]{Wang2021}
Z.~Wang, ``Research on feature and architecture design of ai firewall,'' in
  \emph{5th International Conference on Data Science and Business Analytics
  (ICDSBA)}.\hskip 1em plus 0.5em minus 0.4em\relax IEEE, 2021.

\bibitem[Yokoyama(2019)]{Yokoyama2019}
H.~Yokoyama, ``Machine learning system architectural pattern for improving
  operational stability,'' \emph{IEEE International Conference on Software
  Architecture - Companion (ICSA-C)}, 2019.

\bibitem[Smith(2017)]{Smith2017}
\BIBentryALTinterwordspacing
D.~Smith. (2017) Exploring development patterns in data science – theorylane.
  [Online]. Available:
  \url{https://www.theorylane.com/2017/10/20/some-development-patterns-in-data-science/}
\BIBentrySTDinterwordspacing

\bibitem[Savolainen and Myllärniemi(2009)]{Savolainen2009}
J.~Savolainen and V.~Myllärniemi, ``Layered architecture revisited -
  comparison of research and practice,'' \emph{Joint Working IEEE/IFIP
  Conference on Software Architecture and European Conference on Software
  Architecture (WICSA/ECSA)}, 2009.

\bibitem[Khomh and Gu{\'e}h{\'e}neuc()]{Khomh2020}
F.~Khomh and Y.-G. Gu{\'e}h{\'e}neuc, ``Software engineering patterns for
  machine learning applications (sep4mla).''

\bibitem[Haindl et~al.(2022)Haindl, Buchgeher, Khan, and Moser]{Haindl2022}
P.~Haindl, G.~Buchgeher, M.~Khan, and B.~Moser, ``Towards a reference software
  architecture for human-ai teaming in smart manufacturing,''
  \emph{International Conference on Software Engineering}, 2022.

\bibitem[John and Misra(2017)]{John2017}
\BIBentryALTinterwordspacing
T.~John and P.~Misra. (2017) Lambdaarchitecture pattern. [Online]. Available:
  \url{https://hub.packtpub.com/lambdaarchitecture-pattern/}
\BIBentrySTDinterwordspacing

\bibitem[Ao and of~Engineers(2010)]{Ao2010}
S.~I. Ao and I.~A. of~Engineers, \emph{World Congress on Engineering
  (WCE)}.\hskip 1em plus 0.5em minus 0.4em\relax Newswood Ltd., 2010.

\bibitem[Sculley et~al.(2015)Sculley, Holt, Golovin, Davydov, Phillips, Ebner,
  Chaudhary, Young, Crespo, and Dennison]{Sculley2015}
D.~Sculley, G.~Holt, D.~Golovin, E.~Davydov, T.~Phillips, D.~Ebner,
  V.~Chaudhary, M.~Young, J.-F. Crespo, and D.~Dennison, ``Hidden technical
  debt in machine learning systems,'' \emph{Advances in neural information
  processing systems}, vol.~28, 2015.

\bibitem[Renggli et~al.(2019)Renggli, Karlaš, Ding, Liu, Schawinski, Wu, and
  Zhang]{Renggli2019}
C.~Renggli, B.~Karlaš, B.~Ding, F.~Liu, K.~Schawinski, W.~Wu, and C.~Zhang,
  ``Continuous integration of machine learning models with ease.ml/ci: Towards
  a rigorous yet practical treatment,'' \emph{arXiv}, 2019.

\bibitem[Gustafson(2022)]{Gustafson2022}
\BIBentryALTinterwordspacing
S.~Gustafson, ``How to include machine learning into engineering patterns,''
  2022. [Online]. Available:
  \url{https://www.forbes.com/sites/forbestechcouncil/2022/06/06/how-to-include-machine-learning-into-engineering-patterns/}
\BIBentrySTDinterwordspacing

\bibitem[Take et~al.(2021)Take, Alpers, Becker, Schreiber, and
  Oberweis]{Take2021}
M.~Take, S.~Alpers, C.~Becker, C.~Schreiber, and A.~Oberweis, ``Software design
  patterns for ai-systems.'' in \emph{EMISA}, 2021, pp. 30--35.

\bibitem[Bogner et~al.(2019)Bogner, Boceck, Popp, Tschechlov, Wagner, and
  Zimmermann]{Bogner2019-CSEQUDOS}
J.~Bogner, T.~Boceck, M.~Popp, D.~Tschechlov, S.~Wagner, and A.~Zimmermann,
  ``Towards a {{Collaborative Repository}} for the {{Documentation}} of
  {{Service-Based Antipatterns}} and {{Bad Smells}},'' in \emph{2019 {{IEEE
  International Conference}} on {{Software Architecture Companion}}
  ({{ICSA-C}})}.\hskip 1em plus 0.5em minus 0.4em\relax {Hamburg, Germany}:
  {IEEE}, Mar. 2019, pp. 95--101.

\end{thebibliography}

\end{document}